\documentstyle[floats,aps,psfig]{revtex}

\begin{document}
\draft
\twocolumn[\hsize\textwidth\columnwidth\hsize\csname @twocolumnfalse\endcsname
\title{Haldane Gap and Fractional Oscillations in Gated Josephson Arrays}
\author{Ehud Altman$^1$ and Assa Auerbach$^{1,2}$}
\address{$^1$Department of Physics, Technion, Haifa 32000, Israel.\\
$^{2}$Department of Physics, Stanford University, CA 94305, USA.}
\date{\today }
\maketitle
\begin{abstract}
An analogy between the twisted quantum
$xxz$ model and a gated Josephson junction array 
is used to predict sharp structure in
the critical currents versus gate voltage, and fractional AC Josephson frequencies.
We prove selection rules
for level crossings which  imply fractional periodicities of
ground states with varying Aharonov-Bohm flux. Extrapolated numerical diagonalization
on ladders,   find 
a Haldane gap at moderate easy-plane anisotropy, with vanishing superfluid stiffness.
Physical parameters for  experimental realization of these novel effects are proposed.
\end{abstract}
\vskip2pc]
\narrowtext
Quantum phase fluctuations in superconductors can drive
zero temperature superconductor to insulator 
transitions\cite{doniach}, as observed for example  in disordered films\cite{M-I}.
Their effects can be enhanced and studied in detail using
a weakly coupled, low capacitance Josephson junction array (JJA)\cite{DCHHC,ooden}.

Theory of quantum phase fluctuations
has used models of interacting bosons on a lattice\cite{BHM},
and 
quantum dynamics of vortices\cite{vortices}.  The latter approaches have
proposed  collective phases such as 
vortex Bose condensation\cite{Fisher-Lee}
(for the insulator), and fractional Quantum Hall phases (for JJAs in magnetic field\cite{adi}).

Lattice bosons  map onto effective models of
quantum spins.  A popular approximation to the phase diagram is mean field theory on
the classical (large $S$) spin model\cite{BHM-xxz}. In the strongly quantum regime,
the same mapping relates the  Mott insulator (integer bosons per site)
and  the quantum disordered antiferromagnet\cite{BHM}.

In this paper we explore the quantum magnetism analogy further. We focus our attention
to the effects of a {\em periodic}
lattice  on
superconductivity.
We study the quantum xxz model with twisted boundary
conditions (i.e. an Aharonov Bohm (AB) flux)  
both numerically and analytically. The many body spectrum,
vortex tunneling rates, 
and superfluid stiffness are computed for different lattice dimensions and magnetization 
(Cooper pair density).
We prove general selection rules  for symmetry protected level crossings.
This
rule 
imposes {\em fractional} periodicities of the ground state
as a function  of
AB flux, and is closely related to the ``fractionally quantized phases'' found by Oshikawa, Yamanaka and Affleck
(OYA) in magnetized Heisenberg chains\cite{OYA}.

The following effects may be observed in JJAs of dimensions $L_x\times L_y$.  
\newline
(i) {\em Haldane gap}. Our numerical results for the  $xxz$ model on a two leg ladder
find
a Haldane gap in a sizeable regime of the easy plane anisotropy parameter.
The lowest gap remains finite, while
the superfluid stiffness 
decays exponentially for $L_x \to \infty$. The Haldane phase  
is characterized by a suppressed  critical current (relative to its classical value), and a 
high AC Josephson frequency $f_Q= 2eV/h$,  where the classical  Josephson frequency
of the array is
$f_{cl}= 2eV/(hL_x)$.
\newline
(ii)  {\em Fractional Oscillations.} At fractional  Cooper pair densities $n=p/(q L_y)$, $p,q$ integers,
selection rules derived below  
produce sharp dips in the critical  current $I_{cr}$ versus gate potential.
In these states,  AC Josephson
oscillations appear at {\em subharmonic} frequencies $f_Q^q~=f_Q/q$.  

We conclude by proposing physical parameters for experiments.

The short range Bose Hubbard model is given by
\begin{equation}
H^{BH} = U\sum_i n_i^2   +  \sum_{\langle ij\rangle } \left( V n_i n_j   -2J\left( e^{i\theta_{ij}} b^\dagger_i b_j +
 \mbox{H.c} \right) \right),
\label{BHM}
\end{equation}
where $b^\dagger_i$ creates a  boson  (Cooper pair\cite{comm-BHM}) at site  $i$ 
on a square lattice with nearest neighbor bonds $\langle ij \rangle$, and  $n_i=b^\dagger_i b_i$. 
The
lattice is placed on a cylinder penetrated by an AB flux $\Phi$, introduced via
the gauge phases $\theta_{i  j}=\delta_{j,i+{\hat x}}  2\pi \phi /L_x$, where  $\phi=\Phi/ \Phi_0$, 
and $\Phi_0=  h/(2e c)$.
The supercurrent in the $x$ direction is given by  
$I_s = {1\over h}  \langle \frac{\partial H}{\partial \phi }\rangle$.

At large $U>>J,V$ one can keep  the two lowest energy Fock states
at every site, say $|{\bar n}_i\rangle$ and  $|{\bar n}_i+1\rangle$, and project out all other
occupations.  In the projected subspace,
$b_i^\dagger,b_i, n_i-{\bar n}_i$ 
are replaced by spin half operators $S_i^+, S_i^-$, and $ S_i^z+{1\over 2}$ respectively.
This transformation maps   (\ref{BHM}) onto the quantum $S={1\over 2}$  $xxz$ Model\cite{comm-x-xz}
\begin{equation}
H^{xxz} =\sum\limits_{\langle ij\rangle } \left({J^z \over S^2} S_i^zS_j^z +{J\over 2S^2}   
\left( e^{i\theta _{ij}}S_i^{+}S_j^{-}+\mbox{H.c}\right) \right)
 \label{xxz}
\end{equation}
where the Ising coupling is $J^z= V/4$, and we limit our discussion to easy plane anisotropy $J^z \le J$, in
order to avoid the charge density wave phases\cite{BHM-xxz}.
A pure gauge transformation  $\phi \to \phi+1$ on (\ref{BHM}) or (\ref{xxz}) leaves their spectrum invariant.

\begin{figure}[htb]
\centerline{\psfig{figure=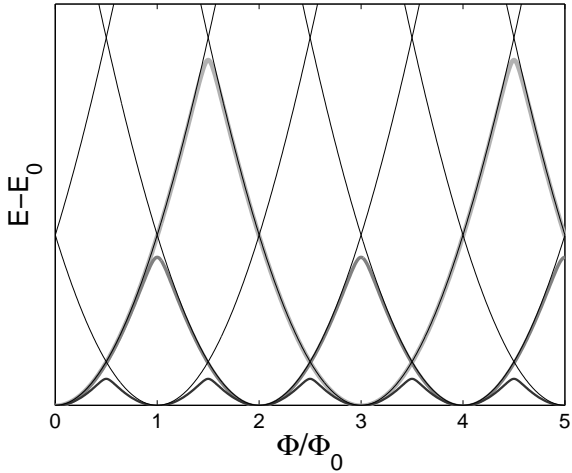,width=3in}}
\vspace{0.5in}
\caption{
\label{fig:adiabats}
Schematic adiabatic ground state energies  as a function of Aharonov Bohm flux $E(\Phi)$, where $E_0=E(0)$.
Thin lines:
classical energies $E_{cl}^i$ (\protect\ref{eqn:Hclass}). Thick lines: quantum adiabats of periods $q\Phi_0$, 
 for $q=1,2,3$.  
Notice the level crossings for $q\ne 1$, which are protected by the selection rules  (\protect\ref{sel-rule}). 
}
\end{figure}

It is instructive to consider the classical (large $S$) ground state energies of  (\ref{xxz}) 
which are adiabatically
connected to the ground states at $\phi=i$, $i=0,1,\ldots$,
\begin{equation}
E^i_{cl}(\phi)=J L_y L_x  \delta n (1- \delta n) \cos\left({2\pi (\phi-i)\over L_x}\right), 
\label{eqn:Hclass}
\end{equation}
where $\delta n\equiv n-{\bar n}$.
At $\phi=q/2$, $q$ integer, pairs of  classical ground states  of oppositely directed supercurrents 
become degenerate, as depicted in  Fig.~(\ref{fig:adiabats}).

Tunneling paths between the two classical ground states can be constructed as
histories of  vortices traversing the lattice in the $y$ direction,
or nucleation and separation of a vortex-antivortex pairs upto the two edges.
The tunneling matrix elements, {\em unless  
prohibited by selection rules,}  open minigaps at the avoided level crossings.
This allows their precise computation from the (many body) eigenenergies as a function
of AB  flux \cite{TH,GT,AA-qhe}.    

The Hamiltonian   (\ref{xxz}) was diagonalized   using the Lanczos algorithm using 
lattice momentum and total magnetization to
block-diagonalize the  matrix. For the two leg ladder at $S^z_{tot}=0$,
we find a regime of $J^z<J$ where the minigaps at $\phi=1/2$
remain finite as $L_x$ increases, which indicates
{\em a Haldane gap phase} in the thermodynamic limit.
This phase has been previously established for
isotropic integer spin chains\cite{haldane},
and half odd integer ladders\cite{ladders} and chains at finite 
magnetic fields\cite{OYA}. 
\begin{figure}[htb]
\centerline{\psfig{figure=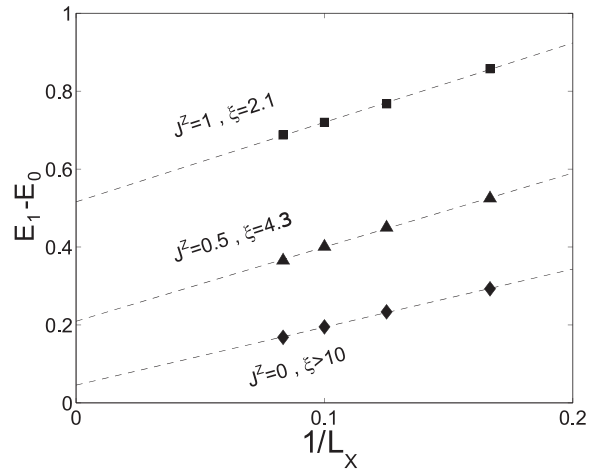,width=3in}}
\vspace{0.5in}
\caption{
\label{fig:haldane}
{\em Haldane gaps}.
Magnon excitation energies  for two leg Josephson ladders for different anisotropy $J^z/J$.
Here $J=1$ and  $E_{m}$ is the lowest eigenenergy of
magnetization  $m$ at flux  $\Phi=\Phi_0/2$.
$\xi$ is the stiffness correlation length given by finite size scaling of ladder  lengths upto  $L_x=12$.
}
\end{figure}

To establish this phase we plot in Fig.~\ref{fig:haldane}, 
the {\em magnon} gap\cite{comm-magnon} at $\phi=1/2$ as a function of $1/L_x$, and extrapolate
the results to $1/L_x \to 0$. However, for the extrapoltion to be justified, we must be certain that
we have reached the asymptotic $L_x >> \xi$ regime where $\xi$ is the correlation length.
$\xi$ was calculated from the superfluid stiffness $K=  \partial^2 E  /\partial \phi^2$,
which was found to fit to
an exponential $K(L_x) \propto \exp(-L_x/\xi)$.  We find that in the regime
$J^z/J^x\in (0.5,1)$, the correlation length is $\xi \in(4.26,2.13)$  which
is safely smaller than the larger system dimensions.
For the pure $xy$ model at $J^z=0$, however,  $\xi$  reaches our
largest system
size. In this regime therefore,  an extrapolated finite gap at $1/L_x=0$ is not credible.

An easy-plane Haldane phase is explained as follows.
The path integral of an even leg ladder of $S=1/2$  spins, can be  mapped onto  
a classical partition function of  an O(2) relativistic field theory in two dimensions\cite{haldane,Book}. 
Its  temperature $T^{2d}$  scales asymptotically as
$\sim  (S L_y)^{-1}$\cite{sudip}. This suggests that below a certain spin size and ladder width,
a disordered phase with exponentially decaying correlations is possible, which
translates into a finite gap for excitations and vanishing stiffness for long ladders.
Above a critical width $L_y > L^{KT}_y$, correlations should decay as a
power-law with a finite (1d) superfluid density
$ 4\pi^2\rho_s =\lim_{L_x \to \infty}( L_x K) > 0$.

{\em Selection rules for avoided level crossings.}
Vortex tunneling is enabled by the lattice since it breaks continuos translational symmetry.
However, the remaining discrete translational symmetry imposes selection rules which are given 
by the following theorem.
\newline
{\em Theorem:}
For the Hamiltonian (\ref{xxz}),  at  $\phi=q/2$, for integer $q$,
any eigenstate $|S^z_{tot}, k_x,\alpha\rangle$
where $k_x$ is the lattice momentum in the $x$ direction,  and $S^z_{tot}$ is the total
 magnetization,
is at least two-fold degenerate
{\em unless} the following condition is satisfied:
\begin{equation}
{S^z_{tot}\over L_x} +{ k_x\over \pi}+ S L_y=p/q~\mbox{($p,q$ integers)}\label{sel-rule}
\end{equation}
The theorem is similar  to Lieb, Shultz and Mattis  (LSM) theorem\cite{LSM} 
for half odd integer  spin chains,
and its extension to  finite magnetizations by OYA
\cite{OYA}.
Here, however, we prove {\em exact}  
degeneracies of the twisted xxz model on {\em finite} lattices, while previous work
concerned gaplessness  in the thermodynamic limit at zero  external
gauge field.

Before providing the proof, let us review three important classes to which the theorem for $S=1/2$ 
applies.
\begin{enumerate}
\item   Odd ladders. $q=1$,   $k_x=0$, $S^z_{tot}=0$.
The selection rules (\ref{sel-rule}) cannot be satisfied,
implying  exact ground
state degeneracy at $\phi=1/2$. This is closely related to the existence
of gapless excitations in the thermodynamic limit of {\em short range} half odd integer spin chains\cite{LSM}.
\item Even ladders with integer magnetization per rung.  The selection rule is obeyed for $q=1$,
which implies a minigap at the first avoided crossing of the ground states.  If this
 minigap survives the  $L_x\to \infty$ limit, the system is in the Haldane gap phase.
\item Even ladders with rational magnetization per rung. The selection rule is obeyed only
for some $q>1$.
This gives rise to a {\em fractional} AB periodicity of the ground state.
\end{enumerate}

{\em Proof:}
The twist  operator is defined as 
\begin{equation}
\hat O\left( \phi \right) \equiv \exp \left( -i\frac{2\pi }{L_x}\phi \sum_{ 
{\bf r}}S^z\left( {\bf r}\right) x\right) .  \label{eqn:O}
\end{equation}
In addition, an $x$-inversion operator $I_x$ is defined $I_x S^\alpha_{x,y} I_x  =S^\alpha_{-x,y}$. 
For any state $\psi_0=|S^z_{tot},k_x,\alpha\rangle$, we define the ``$q$-conjugated'' state $\psi_q$ as 
\begin{equation}
| \psi _q \rangle =\hat O\left(-q\right) I_x | \psi _0 \rangle
\label{psi1}
\end{equation}
{\em Lemma (degeneracy)}:   
\begin{equation}
\langle \psi_q|H(q/2)|\psi_q \rangle  = \langle \psi_0|H(q/2)|\psi_0\rangle
 \label{degen}
\end{equation}
The Lemma is proven by a direct substitution of (\ref{psi1}) in (\ref{degen})
noting  that $O(\phi)$ is the explicit gauge transformation  on $H$,
\begin{equation}
H\left(\phi  \right) =\hat O\left( -\phi \right) H(0) \hat O\left( \phi
\right),   \label{eqn:H(phi)}
\end{equation}
and using the  identities $I_x O(\phi) I_x = O(-\phi)$, and $[H(0) ,I_x] = 0$. 

The theorem is proved  by showing that $\psi_0$ will be transformed by
``q-conjugation'' into an orthogonal state $ \langle\psi_q|\psi_0\rangle=0$ unless
the selection rule (\ref{sel-rule}) is obeyed. The  unit lattice translation
in the $x$ direction is $T_x$. We make use of the two identitites 
\begin{eqnarray}
T_x {\hat O}(-q) T_x^{-1} &=& \exp\left(  i2\pi q S_{tot}^z /L_x +i 2\pi q S L_y  \right)O ( -q  ),\nonumber\\
I_x T_x I_x &=&  T_x^{-1}.
\end{eqnarray}
The lattice momentum of $ \psi_q $ is given by
\begin{equation}
T_x| \psi_q \rangle=\exp \left(-i \frac{2\pi q }{L_x}S_{tot}^z +i 2\pi q S L_y  +i k_x \right) | \psi_q \rangle
\end{equation}
It follows that $\langle\psi_q|\psi_0\rangle=0$ unless the momentum difference
$ \delta k_x =2\pi(q S_{tot}^z/L_x  - q SL_y)+  2 k_x $ is an integer 
multiple of $2\pi$, which proves selection rules (\ref{sel-rule}). QED.

Translating back into the boson language, we consider the ground state at $\phi=0$, with
$k_x=0$, and excess Bose density $\delta n={p \over qL_y}  $.
The  selection rule  implies that as $\phi$ is increased,  the adiabatic ground state passes
 $q-1$ exact
level crossings 
before reaching the first minigap allowed by  (\ref{sel-rule}).
Hence it is clear that  the ground state adiabatic periodicity in AB flux is  
$q \Phi_0$. 
Fig.~\ref{fig:adiabats} depicts the  ground state evolution for the cases $q=1,2,3$.
The critical current is bounded by
\begin{equation}
I_{cr}(q) \le (J/\hbar) L_y \delta n(1-\delta n) \sin(\frac{\pi}{L_x} q)
\end{equation}
which holds upto $q=L_x/2$ where the bound coincides with the classical critical current
$I_{cr}^{cl}= (J/\hbar) L_y \delta n(1-\delta n)$.

\begin{figure}[htb]
\centerline{\psfig{figure=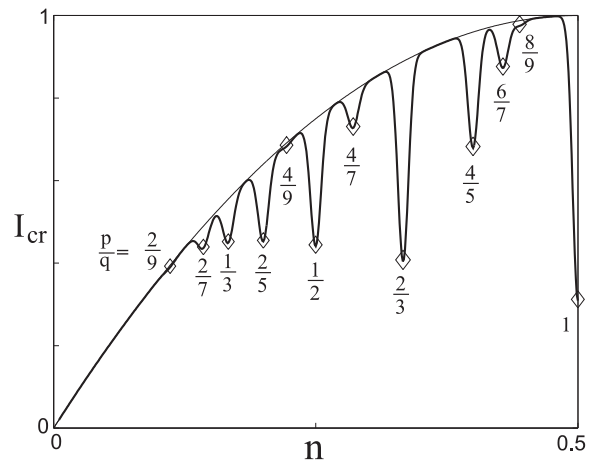,width=3in}}
\vspace{0.5in}
\caption{
\label{fig:Icr}
Schematic diagram of critical currents versus Cooper pair density $n$ for an array of size $L_x$=20
and $L_y$=2. Rational numbers $p/q=n_q L_y $, which label
the dips at $n_q$, are indicated. The classical critical 
current is depicted by a thin line.
Above the critical currents at $n_q$,  fractional AC Josephson frequencies are expected at $2eV/(h q)$.
Spatial disorder and finite temperature are introduced qualitatively by broadening of the dips.
}
\end{figure}

In Fig.~\ref{fig:Icr} the schematic structure of the critical current is plotted 
against the Cooper pairs density for a Josephson ladder with $L_x=20$.
Notice the sharp dips in the critical current at rational densities which obey the  selection
rule. For weak easy plane anisotropy, these minima are  expected to vanish
in the thermodynamic limit, reflecting vanishing stiffness and gapped excitations at these
points.

The classical  Josephson frequency
of an array of length $L_x$  array is
$f_{cl}= 2eV/(hL_x)$, where $V$ is the {\em total} voltage drop in the $x$ direction.
However at  rational densities $n_q=p/(qL_y)$,
for bias current slightly above
$I_{cr}(n_q)$, an AC Josephson effect should be observed with
frequency 
$f_q= 2eV/(qh)$.  This could be pictured as current oscillations caused by moving on the
adiabatic curves of Fig.~\ref{fig:adiabats}.
Alternatively, this effect could perhaps be better detected as fractional  Shapiro steps in
an external high frequency electromagnetic field\cite{SO}.

{\em Experimental realization}.
One of the experimental setups described in  Ref. \cite{ooden}, has  individual gate voltage probes
which control the Cooper pair density at each island to high accuracy. The
short range Bose Hubbard model for this type of JJA can be justified
if the ratio of inter-island capacitance to 
gate capacitance obeys $\epsilon= C/C_0 <<1$.
This translates to
onsite and intersite interactions of Eq. (\ref{BHM}) given by
$U=2e^2(1-4\epsilon)/C_0$, and $V= \epsilon 4e^2 /C_0$ respectively. $U>>V$ in
this regime.
In order map (\ref{BHM}) to (\ref{xxz}),  we demand that $U >>  J$, where  $J$ is 
the Josephson coupling between islands.
The easy-plane regime is given by $V \le 4J$. As emphasized earlier,  this inequality is crucial
for eliminating possible
charge density states, which may also exhibit reduced critical currents at
commensurate fillings.

The junction parameters reported in Ref. \cite{ooden} were
$C_0=$0.64fF , and $C=$1.0fF, and $J=$0.63$^\circ$K. To reach the desired regime, these 
parameters should be modified to approximately
$C_0=$3.0fF, and $C=$0.3fF, and $J=$0.2$^\circ$K.

{\em Acknowlegements}: Useful discussions with J. Avron,  A. Stern, A. van Oudenaarden,
and S-C. Zhang are gratefully acknowledged.  The authors thank the Department of
Physics, Stanford University, where part of this work was performed. AA acknowledges a grant from
the Israel Science
Foundation.


\begin{references}
\bibitem{comm-BHM} While (\ref{BHM}) is 
sometimes used as a coarse grained theory of
homogenous superconductors, it does not contain the fermion (pair breaking) excitations.
\bibitem{comm-x-xz}  For bipartite lattices, we replace the negative Bose hopping
by a positive $xy$ exchange using a sublattice rotation\protect{\cite{Book}}.
\bibitem{comm-magnon} The minigaps of Fig.~\protect{\ref{fig:adiabats}} are between $S^z_{tot}=0$ states,
while magnons of $S^z_{tot}=\pm 1$, have lower excitation energies due to easy plane anisotropy $J^z<J$.
\end{references}
\end{document}